\documentclass[accepted]{article}

\usepackage{microtype}
\usepackage{graphicx}
\usepackage{tabularx}
\usepackage{subfigure}
\usepackage{booktabs} % for professional tables
\usepackage{float}
\usepackage{fancyvrb}

\usepackage{hyperref}

\usepackage{icml2024}

\usepackage{amsmath}
\usepackage{amssymb}
\usepackage{mathtools}
\usepackage{amsthm}

\usepackage[capitalize,noabbrev]{cleveref}

\theoremstyle{plain}

\theoremstyle{definition}

\theoremstyle{remark}

  {\list{}{\leftmargin=0.3in\rightmargin=0.3in}\item[]}%
  {\endlist}

\usepackage{xcolor}
 \newcommand{\WP}[1]{\iftrue
  {\color{magenta} WP: {#1}}
 \fi}

\usepackage[textsize=tiny]{todonotes}

\icmltitlerunning{Using Large Language Models for Humanitarian Frontline Negotiation: Opportunities and Considerations}

\begin{document}

\twocolumn[
\icmltitle{Using Large Language Models for Humanitarian Frontline Negotiation: Opportunities and Considerations}



\icmlsetsymbol{equal}{*}
\begin{icmlauthorlist}
\icmlauthor{Zilin Ma}{equal,yyy}
\icmlauthor{Susannah (Cheng) Su}{equal,yyy}
\icmlauthor{Nathan Zhao}{equal,yyy}
\icmlauthor{Linn Bieske}{yyy}
\icmlauthor{Blake Bullwinkel}{yyy,xxx}
\icmlauthor{Yanyi Zhang}{yyy}
\icmlauthor{Sophia (Yanrui) Yang}{yyy}
\icmlauthor{Ziqing Luo}{yyy}
\icmlauthor{Siyao Li}{yyy}
\icmlauthor{Gekai Liao}{yyy}
\icmlauthor{Boxiang Wang}{yyy}
\icmlauthor{Jinglun Gao}{yyy}
\icmlauthor{Zihan Wen}{yyy}
\icmlauthor{Claude Bruderlein}{zzz}
\icmlauthor{Weiwei Pan}{yyy}

\end{icmlauthorlist}

\icmlaffiliation{yyy}{Harvard School of Engineering and Applied Sciences, Allston, MA, USA}
\icmlaffiliation{zzz}{Harvard T. H. Chan School of Public Health, Cambridge, MA, USA}
\icmlaffiliation{xxx}{Microsoft, Redmond, WA, USA}
%\icmlaffiliation{comp}{Company Name, Location, Country}
%\icmlaffiliation{sch}{School of ZZZ, Institute of WWW, Location, Country}

\icmlcorrespondingauthor{Zilin Ma}{zilinma@g.harvard.edu}
\icmlcorrespondingauthor{Susannah (Cheng) Su}{susannah\_su@college.harvard.edu}
\icmlcorrespondingauthor{Nathan Zhao}{nathanzhao@college.harvard.edu}
\icmlcorrespondingauthor{Weiwei Pan}{weiweipan@g.harvard.edu}

\icmlkeywords{Machine Learning, ICML}

\vskip 0.3in
]



\printAffiliationsAndNotice{\icmlEqualContribution} % otherwise use the standard text.

\begin{abstract}
Humanitarian negotiations in conflict zones, called \emph{frontline negotiation}, are often highly adversarial, complex, and high-risk. Several best-practices have emerged over the years that help negotiators extract insights from large datasets to navigate nuanced and rapidly evolving scenarios.
Recent advances in large language models (LLMs) have sparked interest in the potential for AI to aid decision making in frontline negotiation. Through in-depth interviews with 13 experienced frontline negotiators, we identified their needs for AI-assisted case analysis and creativity support, as well as concerns surrounding confidentiality and model bias. We further explored the potential for AI augmentation of three standard tools used in frontline negotiation planning. We evaluated the quality and stability of our ChatGPT-based negotiation tools in the context of two real cases. Our findings highlight the potential for LLMs to enhance humanitarian negotiations and underscore the need for careful ethical and practical considerations.
\end{abstract}

\section{Introduction}
Humanitarian negotiations in conflict zones, known as \emph{frontline negotiation}, help secure access to crisis-affected populations for aid delivery~\cite{GISF2020}. These negotiations are often highly adversarial, complex, and high-risk. Stakeholders are fragmented geographically, politically, and culturally. Negotiators must \emph{quickly} navigate conflicting perspectives by synthesizing unstructured information, including interviews, stakeholder meetings, and historical documents. Given the time pressure on negotiators, efficient and accurate information synthesis is crucial~\citep[e.g.][]{pon2023, bruderlein2023}.
%Humanitarian negotiations in conflict zones, are called \emph{frontline negotiation}; they are used help to secure access to crisis-affected populations for aid delivery~\cite{GISF2020}. 
%Frontline negotiation are often highly adversarial, complex, and high-risk.
%The stakeholders are often fragmented geographically, politically, and culturally; and negotiators must \emph{quickly} navigate conflicting perspectives by synthesizing unstructured information—including interviews, stakeholder meetings, and historical documents. Given the time pressure on negotiators, efficient and accurate information synthesis is crucial~\citep[e.g.][]{pon2023, bruderlein2023}. 

To address this challenge, negotiators developed templates to assist in manual information processing. Three templates we studied are: the Island of Agreement (IoA), Iceberg and Common Shared Space (CSS), and Stakeholder Mapping (ShM)~\cite{CCHN2019}. Using these templates, negotiators can efficiently organize unstructured data into clear positions and stakeholder intentions. 
Nonetheless, manual processing is time consuming and can introduce errors (which are costly in real-life terms). Here, large language models (LLMs) might be leveraged to assist in information extraction and synthesis for frontline negotiation.

However, LLMs pose risks such as ``hallucinations'' \citep[e.g.][]{aclanthology2024} and susceptibility to privacy attacks \citep[e.g.][]{Neel2023PrivacyII}. Thus, it is unclear if their benefits outweigh the risks for frontline negotiations. This work addresses the following research questions: (1) Can LLMs generate reliable and useful frontline negotiation case summaries? (2) What are additional use cases for LLMs in this domain? (3) What are the specific ethical and practical concerns associated with using LLMs in frontline negotiation? (4) What are the barriers to the adoption of LLM-powered tools by frontline negotiators?

We first explored summarization using OpenAI's GPT-4 API on two real, anonymized frontline negotiation cases to fill out the IoA and ShM templates. We found that GPT-4 showed strong consistency in its responses, producing highly similar summaries (measured by cosine similarity) across 30 API calls for both IoA and Iceberg/CSS. Additionally, GPT-4 generated IoA and Iceberg/CSS results comparable to those manually populated by negotiators. While negotiators typically spend hours or days on these templates, ChatGPT completes them in seconds.

We next conducted interviews with 13 experienced frontline negotiators. Using a video of GPT-4 filling out Iceberg templates with one of the anonymous cases as a design probe \cite{design_probe}, we stimulated discussion with negotiators. We identified two key uses for LLMs in frontline negotiation: (1) \emph{Context Analysis}: Automating parts of context analysis for long documents and unstructured texts. (2) \emph{Ideation Augmentation}: Proposing alternative plans, arguments, or solutions.

%\item \textbf{Support knowledge sharing:} Facilitating training and sharing past cases while addressing anonymization and contextual relevance.
Negotiators also highlighted several concerns about integrating LLMs into their workflow: (1) \emph{Confidentiality:} Proprietary model providers may not ensure data confidentiality. (2) \emph{Western bias in LLMs:} Bias embeddd within LLMs may influence the conclusions drawn by negotiators. (3) \emph{Adoption barriers:} Public, donor and mandator~\footnote{Donors provide financial and material support, while mandators grant the authority and framework for negotiators to represent their interests~\cite{CCHN2019}.} opinion on AI may affect adoption of LLM tools.  (4) \emph{Accuracy and trust:} LLMs are prone to hallucinating and generating ungrounded content. (4) \emph{Incompleteness:} LLM-based automation may fail to capture the emotional and human-centric aspects of frontline negotiation. (5) \emph{Overreliance:} Overreliance may cause users to overlook mistakes made by LLMs.

Overall, our results demonstrate the potential for LLMs to assist frontline negotiators. In particular, we explored auto-filling negotiation templates and generated reliable summaries that are comparable to those written by experts. While many negotiators have already begun using tools like ChatGPT in their daily work, our research is the first to formally study the benefits and risks associated with using LLMs for frontline negotiation. Given the sensitive and high-stakes nature of humanitarian negotiations, we recommend further investigation into the risks we identify to ensure safe and ethical adoption of LLMs in this domain. % we hope that eventually we will have good tools to make negotiation safer by ...

\section{Related Work}
% the current status, progress and challenges
% Check the rehearsal paper from Michael Bernstein 
\subsection{Frontline Negotiation}
Humanitarian negotiators working at the frontlines often engage high-pressure negotiations. In these situations, preparation time is constrained, training opportunities are limited, and the number of stakeholder groups is high. Therefore, building trust and finding common ground to achieve favorable outcomes for those impacted remains a challenge~\cite{CCHN2019, SuttonRhoads2022}. Additionally, exploring all possibilities in a negotiation is difficult. Even in tense hostage situations with armed groups where negotiators have low bargaining power, negotiators rarely fully exploit all of their options, leaving a swath of tactical choices unexplored~\cite{Clements2020}.

\subsection{Templates for Synthesizing Information in Frontline Negotiation}
% TODO: talk about IoA, Iceberg, stakeholder mapping, check CCHN manual. -- addressed by Linn
%\WP{Please reference the appendix sections where more details about these tools are provided!}
%\WP{I think I need an explanation of why it's important to talk about these tools in this paper. That is, we are trying to target the negotiation preparation pipeline through automating these tools}
To address the complication of unstructured information overload in negotiations, practitioners have developed templates to synthesize information in a structured manner, making strategizing easier. We studied three tools in particular: IoA, Iceberg/CSS, and ShM~\cite{CCHN2019} (See also Appendix~\ref{sec:negotiation_tools}).

IoA is a framework that helps negotiators identify convergent and divergent norms and facts, enabling constructive dialogues by determining agreed-upon aspects of a situation \cite{CCHN2019}.

Iceberg/CSS helps negotiators uncover unspoken values, motives, and interests of counterparts. Mapping deeper beliefs identifies shared spaces to inform negotiation strategies \cite{CCHN2019}.

Humanitarian negotiations often involve many dynamic stakeholders. ShM is crucial for understanding their roles and perspectives, helping to identify effective influence paths and prioritize actor engagement \cite{CCHN2019}.

In this work, we assess whether LLMs can populate these tools and be helpful for frontline negotiators. Based on the above background on frontline negotiation and observations of LLM applications in other fields, we propose the following hypotheses:

\textbf{Hypothesis 1:} Provided with detailed information, LLMs can consistently generate stable and reliable outputs across different negotiation frameworks and cases.

\textbf{Hypothesis 2:} LLMs can produce case analyses that are comparable in accuracy to those generated by experienced humanitarian negotiators.

\textbf{Hypothesis 3:} Frontline negotiators are likely to accept and integrate LLM tools into their workflows if these tools prove to be helpful and satisfy key negotiators' needs.

\textbf{Hypothesis 4:} The deployment of LLMs in humanitarian negotiations raises significant ethical and practical concerns, including data confidentiality and model bias.

\section{Method}
\begin{figure}
  \centering
  \includegraphics[width=1\linewidth]{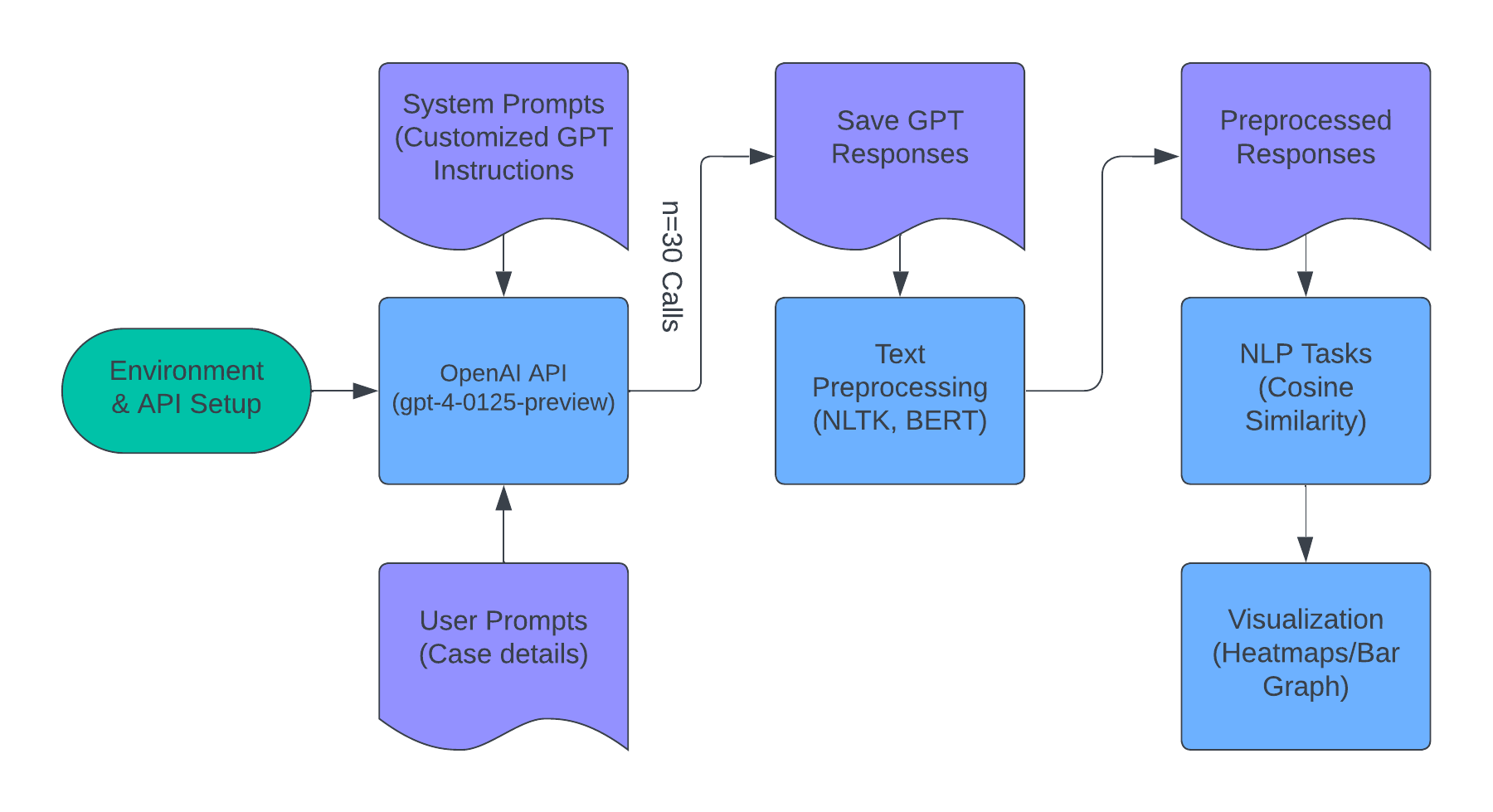}
  \caption{Response Generation and Analysis Pipeline}
  \label{fig:flowchart}
\end{figure}

%\WP{Here I need a high-level description of your pipeline for using GPT to generate these tools before you dive into the details.}
\textbf{Overview: } We configured three Custom GPTs using OpenAI’s GPT-4 to fill in IoA, Iceberg/CSS, and ShM negotiation templates from background information. Users interact through a chatbot interface. The tools were verified for consistency and compared with templates populated by frontline negotiators to assess quality. The evaluation pipeline is shown in Figure~\ref{fig:flowchart}.

However, quantitative results do not necessarily reflect how negotiators integrate tools into their workflow. These metrics are also unable to answer hypotheses \textbf{H3} or \textbf{H4}, which require detailed qualitative feedback from negotiators to address. Beyond context analysis, we are also interested in discovering other potential uses of LLMs in this domain and can only do so by interviewing practitioners directly.
 
\subsection{Quantitative Analysis of LLMs Populating Frontline Negotiation Templates}

\textbf{Tool Development and Iteration:}
% Linn: Elaborate on how the prompt engineering was iterated - compress into one sentence - shortening done
We used an iterative prompt-engineering approach to develop three pipeline tools, starting with system prompts outlining the LLM’s role and expectations. This involved defining task objectives, creating instructions, and specifying output formats, followed by refinement based on feedback from negotiators. The full process is detailed in Appendix~\ref{sec:LLM_tools}

%\WP{Discuss our overall plan for validating the AI tools. Like what are the desirable behaviours we want, how do we quantify these behaviours (metrics) and how do we compute the metrics. Also summarize major findings before diving into the subsections.}
%\WP{Summarize what computational metrics (e.g. stability of response) to make sure that LLM is given ``good" answers.}

% #TODO 2 results: 1. stability of the responses: The built AI-tools deliver robust output when prompted, giving consistant results
%, 2. Benchmarking against negotiators (DONE): The built AI-tools perform similarly well like negotiation practitioners
% 3. finally, mention the time that the negotiators used to generate these reponses. 

\textbf{Evaluating the Stability of the LLM Responses:}
We evaluated the tools based on three principles: stability of responses, benchmarking against negotiator responses, and response time. First, we measured consistency using cosine similarity to ensure low variance and minimal hallucinations. For IoA and Iceberg/CSS, we used \texttt{sklearn.metrics}. For ShM, we used \texttt{sklearn.spatial} to measure the cosine spatial distance between the stakeholder map graph centroids. Second, we benchmarked against human negotiators by calculating semantic similarity between the LLM and human responses for Iceberg/CSS and IoA. Third, we compared the average time ChatGPT takes to generate a response with the average time taken by negotiators, aiming to minimize manual effort for human negotiators.

\textbf{Benchmarking LLM Responses Against Practitioner Responses:}
We benchmarked our LLM-generated responses against those from experienced practitioners. We worked with the Centre of Competence on Humanitarian Negotiation, which shared practitioner-generated respones using the Iceberg and CSS framework and the IoA framework on the ``Health for All'' (HfA) example case. To compare these with the LLM-generated responses, we processed all texts using Bidirectional Encoder Representations from Transformers (BERT) to capture the general semantics more effectively~\cite{devlin2018bert}. After preprocessing with BERT, we conducted the comparison using cosine similarity.

%We chose to employ BERT for preprocessing texts due to its capabilities in understanding and extracting the semantic essence of complex text structures \cite{DBLP:journals/corr/abs-1810-04805}. This is important for accurately grasping the nuances embedded in negotiation-related texts. By preprocessing texts with BERT, we ensured that the subsequent analysis, such as cosine similarity comparisons, was based on deeply contextualized text representations.

\subsection{Human-Centered Benchmarking and Evaluation of Tools}
%\WP{Right now, the two contributions (1) AI tools and (2) qualitative interview, seem really unconnected. Please connect them in terms of goals hypothesis, methodology and insights.}
%\WP{Can you explain specifically how the interviews evaluated the tools? In section 6 and 7 we need to connect section 4 (i.e. tools we made) to the human evaluation.}

After validating the LLM-generated responses using quantitative metrics, we conducted interviews with frontline negotiators to understand their workflow, collect qualitative feedback on the responses, determine whether there are other uses of LLMs in this context, and identify concerns with integrating LLMs into their workflow. 

\textbf{Recruitment: } We disseminated a recruitment survey to the negotiator networks including organizations such as the International Committee of the Red Cross (ICRC), United Nations (UN), and World Food Program (WFP). The survey collected demographic information and years of experience with frontline negotiation. Of the 35 survey responses, 13 interviewees agreed to participate in the interview. We eventually conducted 13 interviews.

\textbf{Semi-structured interview: } We conducted semi-structured interviews with 13 participants, divided into two parts. In the first part, we sought to understand how practitioners prepare for and conduct  negotiations, and the most challenging aspects of their workflows. In the second part, we presented a video demonstration of our Iceberg and CSS tool, developed using a customized GPT. This demonstration served as a design probe~\cite{design_probe} that prompted negotiators to reflect on their workflow, consider other ways in which LLMs might be helpful, and identify weaknesses and risks in the LLM-generated responses. 

Detailed interview guides are provided in Appendix~\ref{apx:interview}. Participants gave informed consent to participate in one hour-long interviews with the option to withdraw and were compensated with a \$30 Amazon Gift Cards. Interview audio recordings were transcribed and anonymized, and the study was approved by our university's Institutional Review Board.

\section{Quantitative Results}

The results of our first experiments, aimed at assessing the stability of ChatGPT's responses, are detailed in Table \ref{tab:cosine_results} in the Appendix. Using 30 API calls, we observed high pair-wise cosine similarity scores across the Food Without Borders (FwB) and Health for All (HfA) scenarios within the Iceberg/CSS and IoA frameworks. For the Iceberg/CSS framework, scores ranged from 0.9627 to 0.9907, with a median of 0.9829 and an average of 0.9821 for FwB, and from 0.9609 to 0.9892, with a median of 0.9767 and an average of 0.9765 for HfA. The IoA framework exhibited scores from 0.9474 to 0.9919, with a median of 0.9788 and an average of 0.9763 for FwB, and from 0.9564 to 0.9895, with a median of 0.9795 and an average of 0.9767 for HfA. The ShM displayed the greatest variability, with scores from 0.7658 to 0.9955, a median of 0.9410, and an average of 0.9265 for FwB, and from 0.7842 to 0.9929, a median of 0.9372, and an average of 0.9231 for HfA. For visual representations of the cosine similarity scores, see Figures \ref{fig:cos_sim_BERT_iceberg_css_fwb}, \ref{fig:cos_sim_BERT_iceberg_css_hfa}, \ref{fig:cos_sim_BERT_IoA_fwb}, \ref{fig:cos_sim_BERT_IoA_hfa}, \ref{fig:cos_sim_BERT_shm_fwb}, and \ref{fig:cos_sim_BERT_shm_hfa} in Appendix \ref{sec:stability}.

%The results of our first set of experiments, aimed at assessing the stability of ChatGPT's responses, are detailed in Table \ref{tab:cosine_results}. These tests involved 30 API calls and demonstrated high pair-wise cosine similarity scores across both the Food Without Borders (FwB) and HfA scenarios within the Iceberg/CSS and IoA frameworks. For the Iceberg/CSS framework, scores ranged between 0.9627 and 0.9907 with a median of 0.9829 and an average of 0.9821 for FwB, and between 0.9609 and 0.9892 with a median of 0.9767 and an average of 0.9765 for HfA. The IoA framework exhibited a broader range, with scores from 0.9474 to 0.9919, a median of 0.9788, and an average of 0.9763 for FwB, and from 0.9564 to 0.9895, a median of 0.9795, and an average of 0.9767 for HfA. The ShM displayed the greatest variability, with scores from 0.7658 to 0.9955, a median of 0.9410, and an average of 0.9265 for FwB, and from 0.7842 to 0.9929, a median of 0.9372, and an average of 0.9231 for HfA. For visual representations of the cosine similarity scores, see Figures \ref{fig:cos_sim_BERT_iceberg_css_fwb} \ref{fig:cos_sim_BERT_iceberg_css_hfa} \ref{fig:cos_sim_BERT_IoA_fwb} \ref{fig:cos_sim_BERT_IoA_hfa} \ref{fig:cos_sim_BERT_shm_fwb} \ref{fig:cos_sim_BERT_shm_hfa} in Appendix \ref{sec:stability}.

The results of our experiments comparing LLM-generated outputs with practitioners' responses for the HfA case are detailed in Figure~\ref{fig:benchmark_ioa_hfa} and Figure~\ref{fig:benchmark_iceberg_css_hfa} in Appendix~\ref{sec:benchmarking_practioners}. In the IoA framework, the average cosine similarity between ChatGPT and practitioner responses was 0.93, with scores ranging from 0.91 to 0.94 across 30 API calls. Similarly, in the Iceberg/CSS framework, the average cosine similarity was 0.92, with scores also ranging from 0.91 to 0.94.

\begin{figure}
  \centering
  \includegraphics[width=1\linewidth]{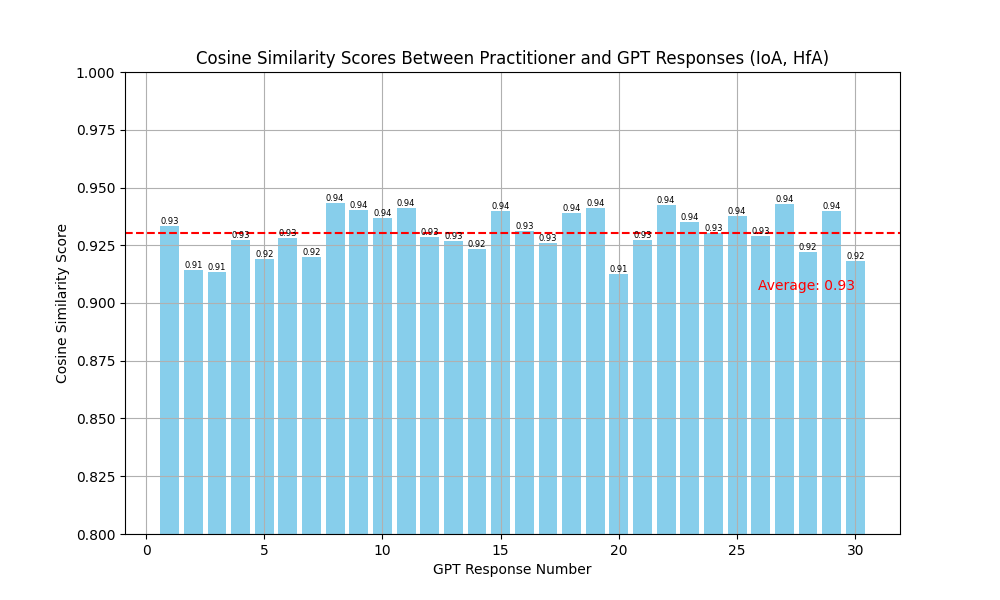}
  \caption{
Cosine similarity scores between practitioner and GPT responses using the IoA framework for the Health for All case averaged \textbf{0.93}. ChatGPT was called 30 times with the same prompt, and responses were compared using BERT. Scores ranged from 0.91 to 0.94, showing consistent alignment with practitioner responses and validating the accuracy of LLM outputs in real-world negotiations.}
  \label{fig:benchmark_ioa_hfa}
\end{figure}

%These benchmarking results, combined with our earlier findings on the stability of LLM-generated responses, provide robust evidence that LLMs are capable of producing accurate, reliable, and consistent outputs that align well with those produced by human experts. This supports Hypothesis 2 that LLMs can be a valuable tool in humanitarian negotiations, offering a high degree of reliability and accuracy.

\section{Interview Results: Opportunities and Concerns of Using LLMs in the Frontline}
%\WP{Can you explain specifically how the interviews evaluated the tools? In section 6 and 7 we need to connect section 4 (i.e. tools we made) to the human evaluation.}

This section summarizes the main findings of our interviews, focusing on the emerging opportunities and concerns regarding the use of LLMs in humanitarian negotiations. Negotiators often referred to LLMs as AI; thus, in the quotes, we retain the term ``AI'' where it was originally used by the participants. Detailed interview analysis is included in Appendix~\ref{sec:opportunities}. 

\subsection{Opportunities}
Although the LLM-generated CSS acted as a design probe and was not meant for practical use, participants appreciated the artifact. For example, P11 mentioned, \textit{``I hope I have this right now.''} Negotiators showed general interest in the next stages of these tools. The CSS tool inspired negotiators to consider ways in which LLMs can be useful in their workflow. Two types of potential novel use cases of LLMs stood out in the interview: 1) context analysis and 2) argument ideation.

\textbf{Context Analysis:} Negotiators face challenges with long, unstructured texts and see potential in LLMs like ChatGPT to automate context analysis and provide support. However, effective prompting is a challenge, and negotiators seek professional guidance to use these tools efficiently, suggesting that LLMs can help them quickly familiarize themselves with cases.

\textbf{Argument Ideation:} Negotiators believe that LLMs can enhance creativity by proposing alternative plans and solutions, helping to avoid narrow-mindedness and find new middle grounds. \textit{``I just don’t want to let my brain be single minded. [...]’’} (P3) Additionally, LLMs can identify information gaps during context analysis, prompting further investigation.

%\textbf{Support knowledge sharing:} Knowledge sharing among negotiators involves training new negotiators and sharing past cases, which is crucial but challenging due to the need for anonymization and contextual relevance. Despite efforts, similar cases often have differing contexts, leading to unsuccessful negotiations. Additionally, the loss of institutional memory hampers the ability to leverage past strategies, suggesting that a decontextualized dataset could be valuable. However, confidentiality agreements limit this potential, indicating that while LLMs could facilitate knowledge sharing and capture institutional memory (supporting Hypothesis 4), ethical and practical concerns must be addressed.

\subsection{Concerns}
Despite positive sentiments for our demo, there are still concerns about using the tools in negotiation.

\textbf{Confidentiality:} Negotiators handle sensitive information and were initially skeptical of using ChatGPT. While comfort with ChatGPT has grown, understanding of relevant privacy policies and appropriate use remains inadequate.

\textbf{Western Bias:} Concerns exist about Western bias in LLMs, affecting their utility in diverse cultural contexts. Feedback from non-Western users is crucial. As one negotiator noted, \textit{``You really need to go to Nigeria, to Myanmar, and talk to the field teams and get their inputs on all these.''} (P12)

\textbf{Public and Mandator Opinions:} Public and mandator views on AI can influence its use in negotiations. Educating the public on AI's limitations and benefits can aid integration into workflows, leading to broader acceptance and more effective use in humanitarian negotiations.

\textbf{Accuracy and Trust:} Concerns about LLM accuracy in negotiation tasks highlight the need for transparency and reliability in LLM-generated responses. LLMs can still serve as a source of inspiration despite these concerns.

\textbf{Limitation of Automation:} LLMs should complement, not replace, the human aspect of negotiation. Human oversight is necessary due to varying levels of preparation and technical skills among negotiators. One negotiator noted, \textit{``The key part of my job is actually to tour the region and engage with different levels [of groups].''} (P8)

\textbf{Overreliance:} There are concerns that over-automating negotiations might hinder effectiveness. Maintaining human involvement in analysis and decision-making is crucial. As emphasized by P4, \textit{``I think it’s important to do your own analysis. I think that information and intelligence comes from doing proper analysis.''}

\section{Discussion}
\textbf{LLMs Generate Consistent and Reliable Templates that Are Comparable to Human Negotiators': } 
% Adopting AI-enhanced negotiation frameworks practitions can speedup their preparation time form ~4 h to only a few minutes 
Our quantitative evaluation of ChatGPT's capabilities in humanitarian negotiations compared AI-generated responses with those from experienced practitioners to assess consistency and reliability. The results showed that LLMs can reliably generate stable, uniform outputs and templates comparable in accuracy to human negotiators. While human negotiators took a week or more to complete templates, ChatGPT did so in seconds. These findings support \textbf{H1} and \textbf{H2}, demonstrating that LLMs can effectively support frontline negotiators by providing consistent templates, thereby enhancing preparation and execution of negotiations.

\textbf{Needs for LLMs in Frontline Negotiation: }
Our qualitative interviews revealed that some negotiators have already started using LLMs for frontline negotiations, primarily for context analysis and argument ideation. These findings support \textbf{H3}, suggesting that frontline negotiators are likely to accept and integrate LLM tools into their workflows if these tools address their key concerns. However, negotiators also emphasized the need for better support in prompt-engineering and a deeper understanding of LLM limitations. Future studies should focus on these areas, and training programs within humanitarian organizations should be designed to help negotiators better understand and effectively use these tools.

%\subsection{Properly Designed LLM tools can Increase Preparedness and Safety}
%Humanitarian negotiator who are operating in conflict zones are often exposed to a high level of risk also threatening their personal safety. The presented LLM tools could help to enhance the training and preparedness of practitioners enabling them to ultimately make better decisions and operate more safely in high-stake situations. Moreover, the AI-enhanced capability to reason about complex and dynamically changing information could enable more robust trust building, identification of influencing strategies and in turn lead to better negotiation outcomes seeking to safeguard most vulnerable groups or conflict resolution.

%While the high similarity scores are promising, it is essential to address ethical and practical concerns, such as data confidentiality and potential biases. Ensuring that AI tools adhere to strict privacy standards and transparent usage policies will be crucial for their acceptance and integration into humanitarian negotiation practices.

\textbf{Concerns over LLMs in Negotiation: }
Privacy and confidentiality are paramount concerns for humanitarian negotiators. Therefore, any LLM system introduced must guarantee stringent privacy standards and clearly disclose its terms. Leakage of confidential information could endanger negotiators and other involved parties, leading to distrust and disruptions. 

Biases are present in every negotiation domain undermining outcomes. LLMs could further skew viewpoints if they provide, for example, Western biased summaries to negotiators operating in other geographies. These findings support \textbf{H4} and reveal a strong need to align LLMs to address these practical and ethical concerns. Since negotiators have already begun integrating LLMs into their workflows, it is crucial to build their capacity to mitigate or identify LLM biases. Consequently, there is a pressing need for comprehensive educational programs and enhanced interface support across organizations to address these limitations.

\textbf{Addressing Privacy and Confidentiality Concerns with Infrastructure Investments: }
Privacy and confidentiality concerns can be partially addressed through the right infrastructure investments. For example, many open-source LLMs \cite{bigscience2022bloom, zhang2022opt, black2022gpt_neox, falcon180b2023} and models optimized for local deployment \cite{abdin2024phi3, google2023gemma, mistral2023mistral, salesforce2023xgen} can generate text comparable to, or even better than, ChatGPT in some domains. By properly observing data and privacy practices, we can ensure the confidentiality required by negotiators. Future work in this area might include comparing the performance of open-source models with our GPT-based negotiation tools.

\textbf{Address Trust, Overreliance and Overautomation before Adaptation: } % should address bias, overreliance, trust, public and mandator's opinion
Concerns about adapting LLMs in negotiation come not only from negotiators themselves but also involved communities, donors, and negotiation mandators. These concerns often stem from broader issues inherent in LLMs that remain unaddressed. We have identified risks related to trust, over-reliance, and over-automation. To successfully integrate LLM tools into negotiation processes, technologists need to demonstrate a thorough understanding of these key issues (e.g., do negotiators trust LLMs even when LLMs make mistakes?) and address them through improved algorithms and interface design.

Technologists must carefully consider where automation is appropriate. Studies have indicated that careless automation only increases manual labor \cite{Beede2020AHE, ulloa_invisible_2022}. For example, negotiators noted that their standard practices do not align with the input formats suited for LLMs. This misalignment affects the adoption of LLM tools and necessitates changes in existing data management practices, such as capturing notes or case data more systematically and continuously. Negotiators might perceive these changes as burdensome. To address this issue, further design work is needed to support negotiators beyond summarization, including aiding in data collection and providing other workflow support.

\subsection{Conclusion}
Our research is the first to demonstrate the potential for LLMs to assist frontline negotiators by auto-filling negotiation templates and providing consistent, reliable summaries that are comparable in quality to those written by human experts. Interviews with negotiators indicated key uses for LLMs in context analysis and ideation augmentation, but also highlighted concerns surrounding confidentiality, embedded biases, barriers to adoption, accuracy, and overreliance. While many negotiators already use tools like ChatGPT, our study formally assesses the benefits and risks, emphasizing the need for further investigation to ensure safe and ethical adoption. Ultimately, we hope that LLMs can enhance the efficiency, safety and effectiveness of humanitarian negotiations, aiding the delivery of critical aid to crisis-affected populations.

\bibliography{example_paper, zilinma}
\bibliographystyle{icml2024}

%%%%%%%%%%%%%%%%%%%%%%%%%%%%%%%%%%%%%%%%%%%%%%%%%%%%%%%%%%%%%%%%%%%%%%%%%%%%%%%
%%%%%%%%%%%%%%%%%%%%%%%%%%%%%%%%%%%%%%%%%%%%%%%%%%%%%%%%%%%%%%%%%%%%%%%%%%%%%%%
% APPENDIX
%%%%%%%%%%%%%%%%%%%%%%%%%%%%%%%%%%%%%%%%%%%%%%%%%%%%%%%%%%%%%%%%%%%%%%%%%%%%%%%
%%%%%%%%%%%%%%%%%%%%%%%%%%%%%%%%%%%%%%%%%%%%%%%%%%%%%%%%%%%%%%%%%%%%%%%%%%%%%%%
\newpage
\appendix
\onecolumn
\section{Appendix}

%%%%%%%%%%%%%%%%%%%%%%%%%
%%%%%%%%%%%%%%%%%%%%%%%%%
%%%%%%% LLM Enhanced Tools for Humanitarian Negotiation
%%%%%%%%%%%%%%%%%%%%%%%%%
%%%%%%%%%%%%%%%%%%%%%%%%%
\subsection{Synthesizing Tool Development}
\label{sec:LLM_tools}

We leveraged a prompt-engineering approach to develop the AI-enhanced negotiation tools: The development of these tools involved a iterative process of creating structured system prompts that guide the LLM in performing the synthesizing task. Each tool begins with a system prompt setup that outlines the role and expectations of the LLM.
\begin{itemize}
    \item IoA Tool: ``You are a highly structured and detail-oriented assistant specializing in negotiation analysis. Based on the provided information about a high-stakes situation, your task is to analyze and summarize the critical information to fill out an Island of Agreement (IoA) table and recommend negotiation strategies.''
    \item Iceberg and CSS Tool: ``You are an assistant in frontline negotiation, specializing in analyzing and summarizing critical information for high-stakes situations like crisis response, emergency management, or conflict resolution. Your task is to help users understand and articulate the positions, reasoning, and motives behind both their organization and the counterparty involved in a negotiation.''
    \item Stakeholder Mapping Tool: ``As the Negotiation Navigator, your role is to assist users in crafting clear and detailed stakeholder mapping visualizations, leveraging provided documents and evolving user inputs. You guide users through a systematic process involving greeting, information extraction, consistent graph design and creation using a standardized Python matplotlib script, adjustments, and analysis of stakeholder dynamics and negotiation paths.''
\end{itemize}

Following the initial setup, the instructions then dive into the specific components of each negotiation tool along with step-by-step instructions and requirements on output formats.

We enhanced our manual configuration instructions through interactions with the ChatGPT web interface and integrated feedback from humanitarian negotiation practitioners to ensure the tools’ practical relevance and effectiveness.

\paragraph{Island of Agreement (IoA)} We give an example of how we instantiated our pipeline to output the Island of Agreement.
We instructed the LLM to populate the IoA table in steps:
\begin{itemize}
    \item Step 1: Sorting and qualifying elements - This initial step involves differentiating between `facts'—such as demographic data, logistical specifics of aid programs—and `norms'—including legal rights and operational priorities. The aim is to categorize these elements based on their relevance and the evidence supporting them.
    \item Step 2: Recognizing areas of promise and challenge - This involves identifying parts of the dialogue that are promising for fostering relationships and those that highlight critical issues needing negotiation.
    \item Step 3: Developing common understanding - The final step focuses on building a mutual understanding with counterparts regarding the negotiation's starting point, emphasizing specific objectives.
\end{itemize}

For each category within the IoA table, the LLM assistant is instructed to provide structured lists of items such as:
\begin{itemize}
    \item Contested Facts: [Fact1, Fact2, ...]
    \item Agreed Facts: [Fact1, Fact2, ...]
    \item Convergent Norms: [Norm1, Norm2, ...]
    \item Divergent Norms: [Norm1, Norm2, ...]
\end{itemize}

Furthermore, based on the analysis, recommendations are formulated on aspects to prioritize and those to avoid during the negotiation:
\begin{itemize}
    \item Prioritize: [Item1, Item2, ...]
    \item Avoid: [Item1, Item2, ...]
\end{itemize}

We also instructed LLM to make the responses concise and structured for easy parsing. We noted in the instruction that the background and detailed case information will be provided by users. For the full instruction we used for IoA, see Appendix \ref{apx:instructions_css}. 

\subsubsection{Iceberg and Common Shared Space (CSS)}
We used a similar approach as what we used in IoA to develop our instructions for Iceberg and CSS. We instructed the LLM to provide the positions, reasoning, motives and values for the counterparty and the user's organization respectively in a concise and easy-for-parsing format. For full instructions, see Appendix \ref{apx:instructions_ioa}.

\subsubsection{Stakeholder Mapping}
We instructed the LLM to analyze documents methodically to ensure stakeholders, their roles, relationships, and the negotiation context are identified consistently, laying a solid foundation for accurate graphing. Emphasis was placed on identifying the protagonist and the counterparty.

Next, we instructed the LLM to use a standardized \texttt{matplotlib} script to draft initial and subsequent stakeholder maps. The script is adaptive and includes the variables identified during the information extraction phase, ensuring a consistent and accurate representation of the negotiation landscape. Emphasis was again placed on proper plotting of the counterparty in red at the center of the graph.

After creation of the graph, the tool will to offer recommendations to the user in identifying efficient negotiation paths, employing network analysis and shortest path algorithms to suggest strategic approaches. For the full set of instructions, see Appendix \ref{apx:instructions_shm}.

%%%%%%%%%%%%%%%%%%%%%%%%%
%%%%%%%%%%%%%%%%%%%%%%%%%
%%%%%%% Tools for Frontline Humanitarian Negotiation Planning
%%%%%%%%%%%%%%%%%%%%%%%%%
%%%%%%%%%%%%%%%%%%%%%%%%%
\subsection{Tools for Frontline Humanitarian Negotiation Planning}
\label{sec:negotiation_tools}
\subsection{Island of Agreement (IoA)}
The objective of IoA in negotiation is to systematically identify and categorize areas of agreement and disagreement between negotiating parties to facilitate productive dialogue and effective negotiation strategies \cite{CCHN2019}.

The IoA table is divided into four distinct categories:

\begin{itemize}
    \item Contested Facts - Facts requiring further evidence for clarification.
    \item Agreed Facts - Established facts that form the basis for initiating dialogue.
    \item Convergent Norms - Norms where parties' values align and can be emphasized as common ground.
    \item Divergent Norms - Areas of normative disagreement that require strategic negotiation.
\end{itemize}

\subsection{Iceberg and Common Shared Space (CSS)}
\begin{figure}[H]
  \centering
  \includegraphics[width=0.6\linewidth]{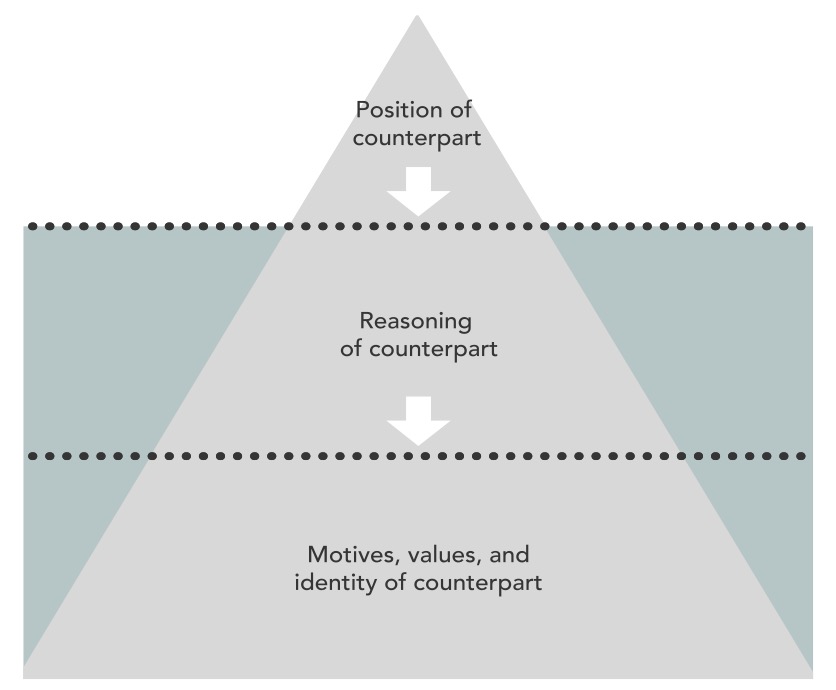}
  \caption{Graphic Example of Iceberg and CSS Framework \cite{CCHN2019}}
  \label{fig:iceberg}
\end{figure}

The Iceberg and CSS framework helps uncover both the explicit demands and positions and also the underlying reasoning and values that influence these stances in humanitarian negotiation \cite{CCHN2019}. It divides the negotiation analysis into three connected layers, each representing a different depth of interaction and understanding, hence the ``iceberg'' \ref{fig:iceberg}:
\begin{itemize}
    \item Positions (What): This is the visible tip of the iceberg. It represents the explicit demands or stances of both parties. It involves identifying and articulating the explicit positions of each party involved in the negotiation.
    \item Reasoning (How): This layer is beneath the surface of the ocean. It explores tactical reasoning that underpins these positions, providing insights into the strategies and logical constructs that lead to the formation of these stances.
    \item Motives and Values (Why): This is the deepest layer. It discovers the core motives and values driving the parties' reasoning and positions. This layer is crucial for identifying potential areas of convergence or divergence that could be leveraged to facilitate dialogue and compromise.
\end{itemize}

\subsubsection{Stakeholder Mapping}
Stakeholder mapping is designed to systematically identify and analyze the influence and interrelations of diverse actors within complex negotiation environments. This tool categorizes stakeholders along two principal axes: their positions on key issues (transformative to conservative) and their levels of influence (global to local, or other customized standards). By visualizing these dimensions, negotiators can strategically assess where stakeholders stand regarding the negotiation's goals, and identify potential allies and adversaries \cite{CCHN2019}. 

\begin{figure}[H]
  \centering
  \includegraphics[width=0.6\linewidth]{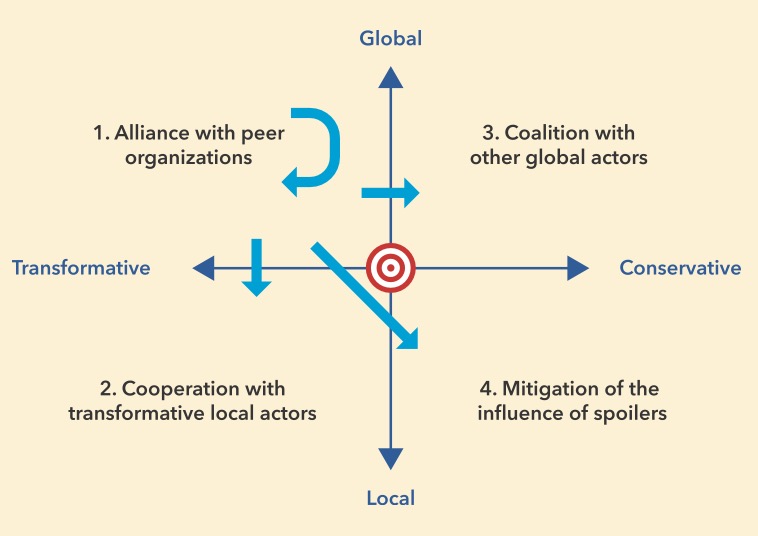}
  \caption{Example of Stakeholder Mapping \cite{CCHN2019}}
  \label{fig:stakeholder}
\end{figure}

%%%%%%%%%%%%%%%%%%%%%%%%%
%%%%%%%%%%%%%%%%%%%%%%%%%
%%%%%%% Stability of LLM-Generated Responses
%%%%%%%%%%%%%%%%%%%%%%%%%
%%%%%%%%%%%%%%%%%%%%%%%%%
\subsection{Stability of LLM-Generated Responses}
\label{sec:stability}
% TODO: the comparison for the negotiator and the AI generated results. tables and numbers
After achieving satisfactory outcomes through the ChatGPT web interface in designing our negotiation tools, we translated our instructions for Custom GPTs into system prompts and case information into user prompts for OpenAI API calls (model: gpt-4-0125-preview). We submitted identical prompts to each customized tool 30 times to evaluate their consistency and reliability.

Similar to the methodology mentioned in the benchmarking step (section 5.4), we processed all texts using BERT. Pairwise comparisons of the LLM-generated responses were conducted to verify consistency.

Cosine similarity heatmaps visualized the stability of responses over 30 calls for each case scenario, showing the pairwise cosine similarity scores between all generated responses and highlighting the model's reliability.

\begin{figure}[H]
  \centering
  \includegraphics[width=1\linewidth]{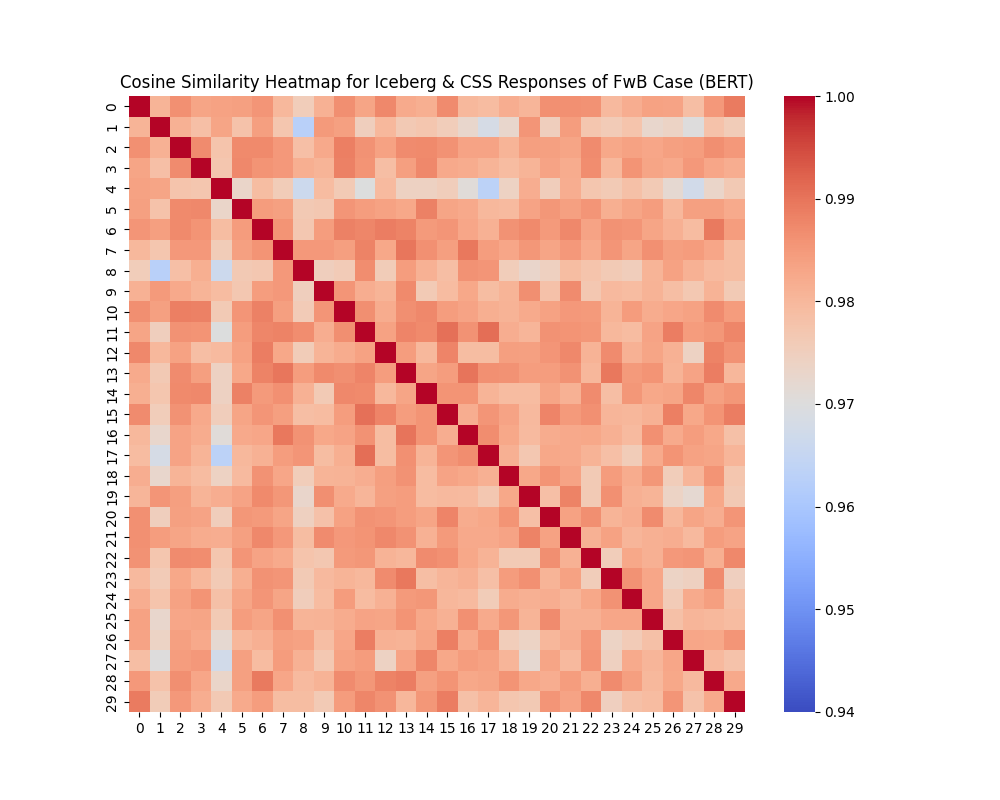}
  \caption{Cosine Similarity of Iceberg and CSS on FwB Case\\This heatmap visualizes the pairwise cosine similarity scores between ChatGPT responses for the Iceberg/CSS framework-based (FwB) case. The texts were processed using BERT, and the heatmap highlights high consistency with median similarity scores above 0.98, indicating stable and reliable outputs across 30 calls.}
  \label{fig:cos_sim_BERT_iceberg_css_fwb}
\end{figure}

\begin{figure}[H]
  \centering
  \includegraphics[width=1\linewidth]{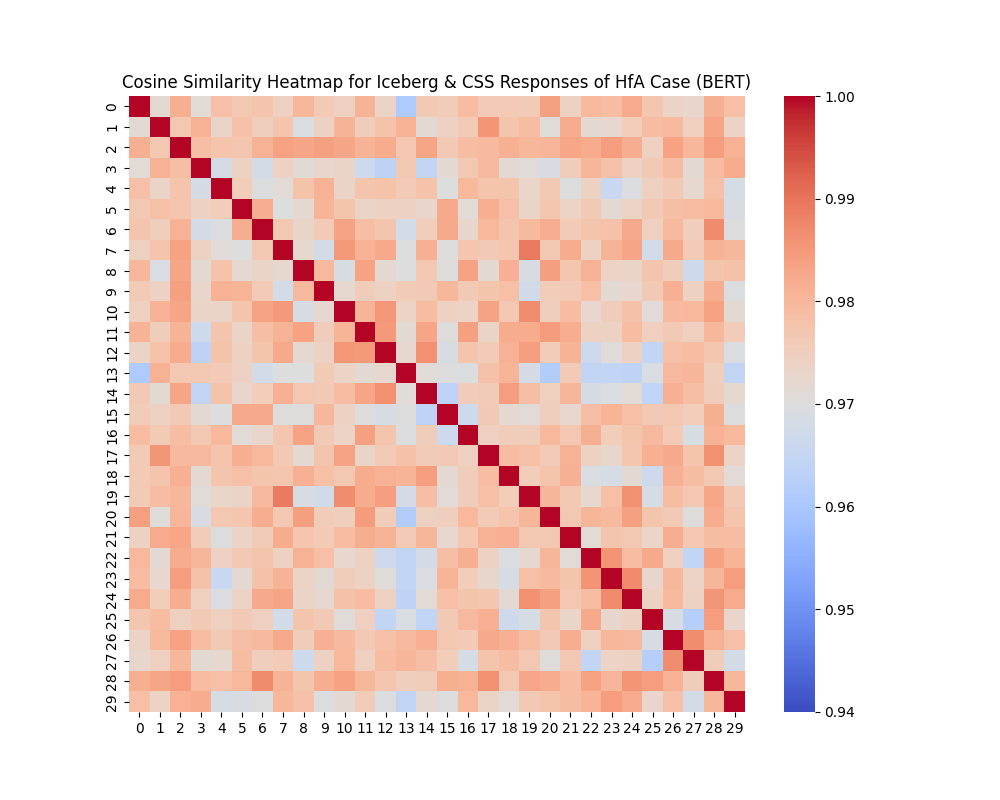}
  \caption{Cosine Similarity of Iceberg and CSS on HfA Case\\This heatmap visualizes the pairwise cosine similarity scores between ChatGPT responses for the Iceberg/CSS framework-based (HfA) case. The texts were processed using BERT, and the heatmap highlights high consistency with median similarity scores above 0.97, indicating stable and reliable outputs across 30 calls.}
  \label{fig:cos_sim_BERT_iceberg_css_hfa}
\end{figure}

\begin{figure}[H]
  \centering
  \includegraphics[width=1\linewidth]{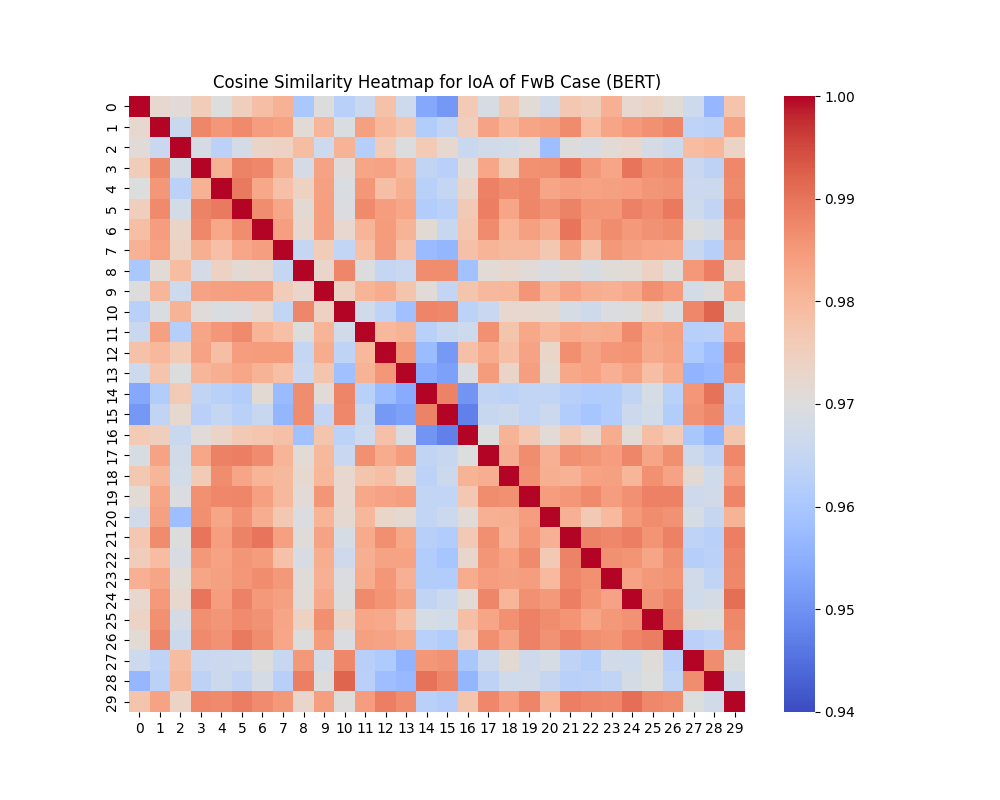}
  \caption{Cosine Similarity of IoA on FwB Case\\This heatmap visualizes the pairwise cosine similarity scores between ChatGPT responses for the IoA framework-based (FwB) case. The texts were processed using BERT, and the heatmap highlights high consistency with median similarity scores above 0.97, indicating stable and reliable outputs across 30 calls.}
  \label{fig:cos_sim_BERT_IoA_fwb}
\end{figure}

\begin{figure}[H]
  \centering
  \includegraphics[width=1\linewidth]{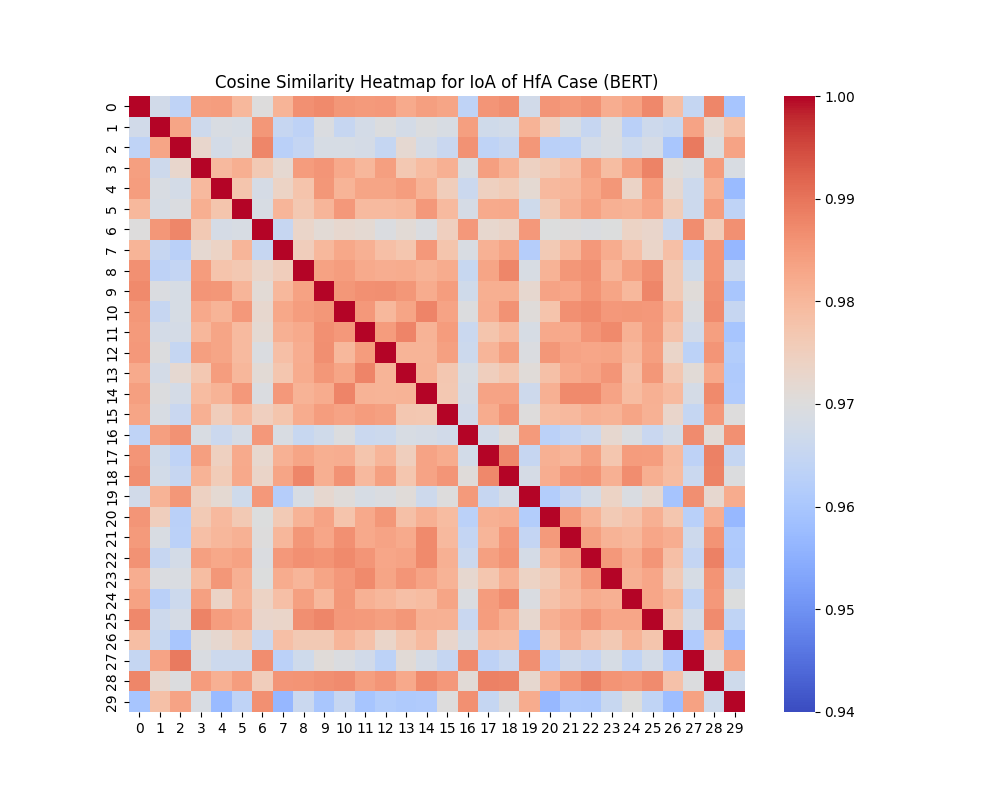}
  \caption{Cosine Similarity of IoA on HfA Case\\This heatmap visualizes the pairwise cosine similarity scores between ChatGPT responses for the IoA framework-based (HfA) case. The texts were processed using BERT, and the heatmap highlights high consistency with median similarity scores above 0.97, indicating stable and reliable outputs across 30 calls.}
  \label{fig:cos_sim_BERT_IoA_hfa}
\end{figure}

The heatmaps \ref{fig:cos_sim_BERT_iceberg_css_fwb} \ref{fig:cos_sim_BERT_iceberg_css_hfa} \ref{fig:cos_sim_BERT_IoA_fwb} \ref{fig:cos_sim_BERT_IoA_hfa} and summary statistics \ref{tab:cosine_results} indicate high degrees of similarity across responses for the Iceberg/CSS and IoA tools, with median cosine similarities consistently above 0.97 for most scenarios. These results suggest a high level of stability in the LLM outputs, supporting the model's reliability in generating consistent responses.

\begin{figure}[H]
  \centering
  \includegraphics[width=0.85\linewidth]{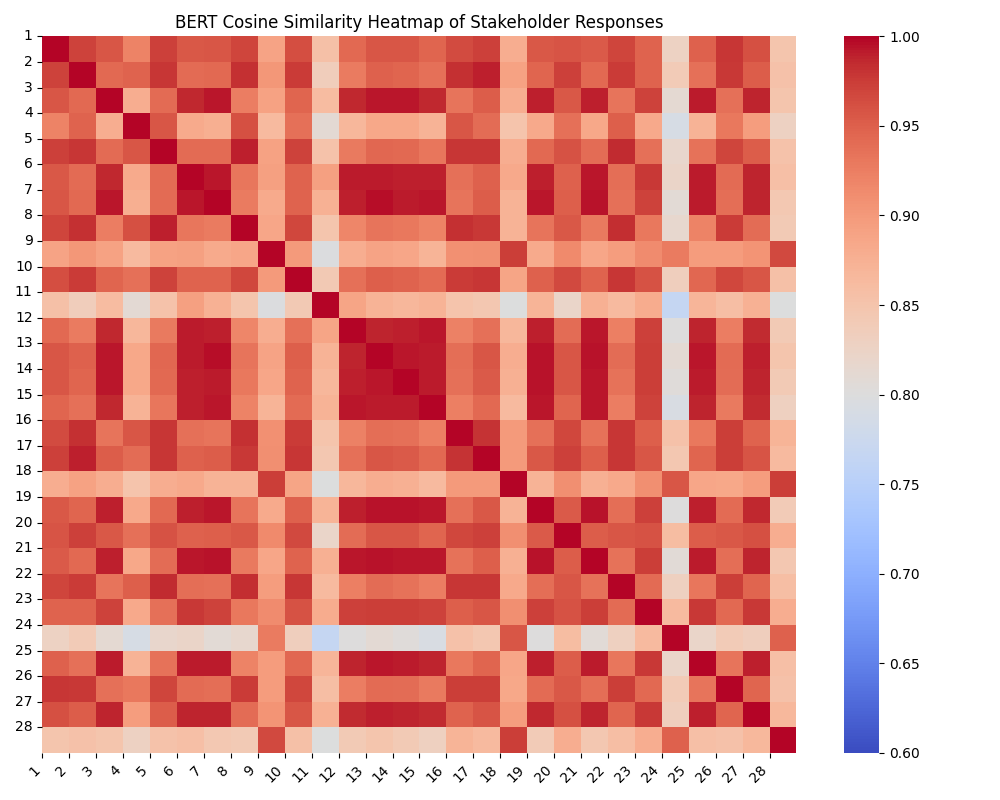}
  \caption{Cosine Similarity of Stakeholder Mapping on FwB Case* \\
  *Note the scale on Figures \ref{fig:cos_sim_BERT_shm_fwb} and \ref{fig:cos_sim_BERT_shm_hfa} (0.6 to 1) is far wider than on Figures \ref{fig:cos_sim_BERT_iceberg_css_fwb}, \ref{fig:cos_sim_BERT_iceberg_css_hfa}, \ref{fig:cos_sim_BERT_IoA_fwb}, \ref{fig:cos_sim_BERT_IoA_hfa} (0.94 to 1).\\This heatmap visualizes the pairwise cosine similarity scores between ChatGPT responses for the Stakeholder Mapping framework-based (FwB) case. The texts were processed using BERT, and the heatmap highlights high consistency with median similarity scores above 0.94, indicating stable and reliable outputs across 30 calls.}
  \label{fig:cos_sim_BERT_shm_fwb}
\end{figure}

\begin{figure}[H]
  \centering
  \includegraphics[width=0.85\linewidth]{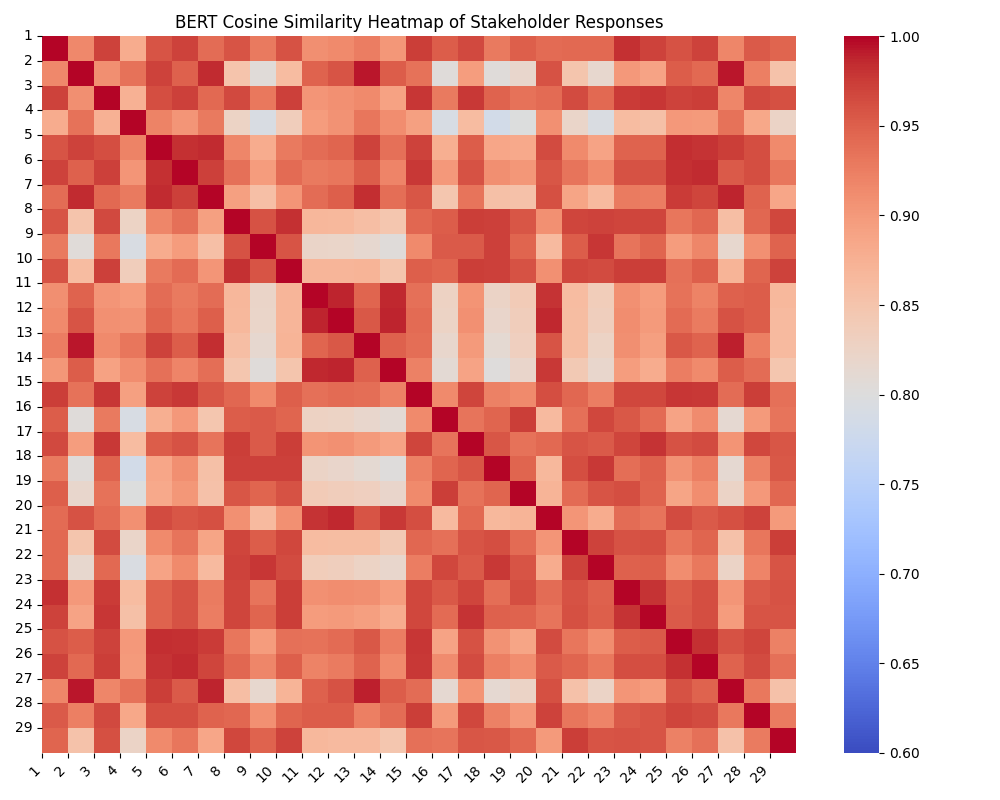}
  \caption{Cosine Similarity of Stakeholder Mapping on HfA Case*\\This heatmap visualizes the pairwise cosine similarity scores between ChatGPT responses for the Stakeholder Mapping framework-based (HfA) case. The texts were processed using BERT, and the heatmap highlights high consistency with median similarity scores above 0.93, indicating stable and reliable outputs across 30 calls.}
  \label{fig:cos_sim_BERT_shm_hfa}
\end{figure}

The heatmaps \ref{fig:cos_sim_BERT_shm_fwb} and \ref{fig:cos_sim_BERT_shm_hfa} indicate slightly lower degrees of similarity for the Stakeholder Mapping tool, with median centroid cosine similarities of about 0.94 for both FwB and HfA cases. These results still suggest a high level of stability, but with notable and outlier exemptions. For example, in contrast to IoA and Iceberg/CSS, the minimum centroid cosine similarity value is significantly lower at $\approx 0.76-0.78$. 

These computational validation results support Hypothesis 1, demonstrating that provided with detailed information, LLMs (ChatGPT) can consistently generate stable and reliable outputs across different negotiation frameworks and cases. The high similarity scores across repeated trials highlight the model's ability to produce reliable and uniform outputs, essential for practical applications in humanitarian negotiations.

\begin{table}[h]
\centering
\caption{Summary of Cosine Similarity Results of ChatGPT Responses\\This table summarizes the cosine similarity scores for various frameworks and cases. It includes the minimum, maximum, median, and average similarity scores for pair-wise comparisons between ChatGPT responses across different frameworks: Iceberg/CSS, IoA, and ShM. The text was processed with BERT before comparison to capture the semantic essence. The scores highlight ChatGPT's response consistency, with higher average scores indicating greater consistency in generating similar responses across multiple calls.}
\label{tab:cosine_results}
\begin{tabularx}{\columnwidth}{@{}l*{4}{>{\centering\arraybackslash}X}@{}}
\toprule
\textbf{Metric} & \textbf{Min} & \textbf{Max} & \textbf{Median} & \textbf{Avg} \\ \midrule
\textbf{Iceberg/CSS FwB} & 0.9627 & 0.9907 & 0.9829 & 0.9821 \\
\textbf{Iceberg/CSS HfA} & 0.9609 & 0.9892 & 0.9767 & 0.9765 \\
\textbf{IoA FwB}            & 0.9474 & 0.9919 & 0.9788 & 0.9763 \\
\textbf{IoA HfA}            & 0.9564 & 0.9895 & 0.9795 & 0.9767 \\
\textbf{ShM FwB}         & 0.7658 & 0.9955 & 0.9410 & 0.9265 \\
\textbf{ShM HfA}         & 0.7842 & 0.9929 & 0.9372 & 0.9231 \\
\bottomrule
\end{tabularx}
\end{table}

%%%%%%%%%%%%%%%%%%%%%%%%%
%%%%%%%%%%%%%%%%%%%%%%%%%
%%%%%%%  Benchmarking LLM Responses Against Practitioner Responses
%%%%%%%%%%%%%%%%%%%%%%%%%
%%%%%%%%%%%%%%%%%%%%%%%%%
\subsection{Benchmarking LLM Responses Against Practitioner Responses}
\label{sec:benchmarking_practioners}
\begin{figure}[H]
  \centering
  \includegraphics[width=1\linewidth]{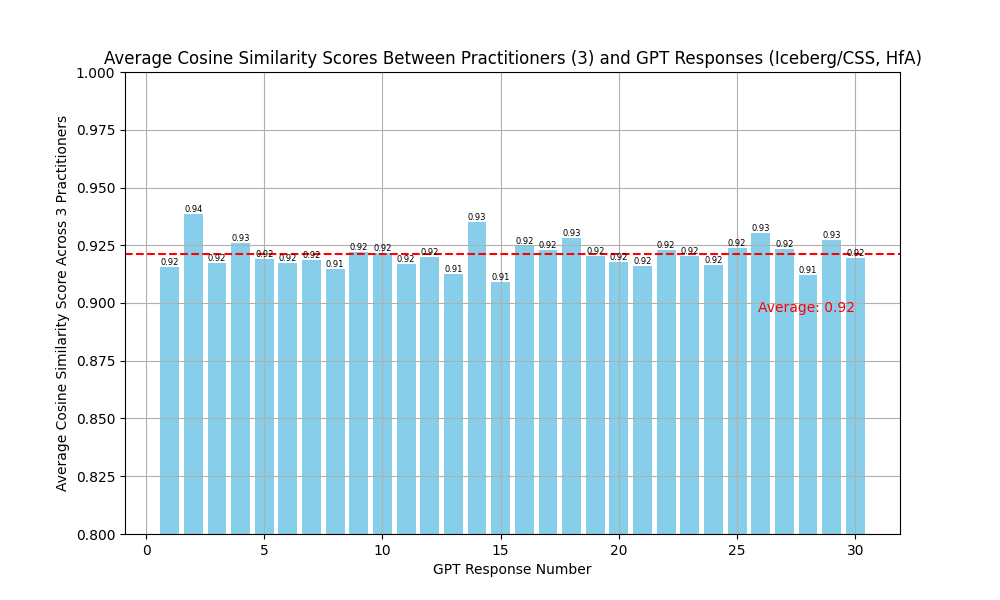}
  \caption{Cosine Similarity Scores Between Three Practitioners' responses and GPT Responses (Iceberg/CSS framework for analyzing the Health for All case). Average Cosine Similarity achieved is \textbf{0.92}.\\ Using the same prompt, ChatGPT was called 30 times to generate responses. Each ChatGPT response was compared to each of the three practitioner responses, and the average similarity score was calculated. The comparison was conducted using BERT to preprocess texts and capture their semantic essence. The scores range from 0.91 to 0.94. It shows LLM-generated responses consistently match practitioner responses across multiple calls, validating the practical relevance and accuracy of LLM outputs in real-world negotiation contexts.}
  \label{fig:benchmark_iceberg_css_hfa}
\end{figure}

%%%%%%%%%%%%%%%%%%%%%%%%%
%%%%%%%%%%%%%%%%%%%%%%%%%
%%%%%%%  GPT-4 Prompts
%%%%%%%%%%%%%%%%%%%%%%%%%
%%%%%%%%%%%%%%%%%%%%%%%%%
\subsection{GPT-4 Prompts}
\label{sec:gpt-4_prompts}

\subsubsection{Iceberg/CSS}
\label{apx:instructions_css}
You are an assistant in frontline negotiation, specializing in analyzing and summarizing critical information for high-stakes situations like crisis response, emergency management, or conflict resolution. Your task is to help users understand and articulate the positions, reasoning, and motives behind both their organization and the counterparty involved in a negotiation. You guide users to provide background information and details about the case, then use the Iceberg and Common Shared Space (CSS) models to frame the negotiation landscape.

Given the provided context, counterparty, and organization, along with three aspects that need to be addressed, summarize the position, reasoning, and motive and values for both parties. 

Organize this information into bullet points under each aspect, adhering to the following logic:
For the Counterparty:
Positions (What): Articulate the counterparty's stance or demands.
Reasoning (How): Assess the tactical reasoning that supports their positions.
Motives and Values (Why): Identify the underlying motives and values driving their reasoning.
For the User’s Organization:
Motives and Values (Why): Start with the organization's core motives and values relevant to the negotiation.
Reasoning (How): Describe the logical or strategic reasoning stemming from these motives/values.
Positions (What): Conclude with the organization's positions or demands based on the reasoning.

Your responses should be coherent with the provided context, aiding users in navigating the negotiation process effectively. You communicate in a formal and professional tone, avoiding flashy language, to mirror the communication style of negotiators.

Please provide the response in a concise, structured format that can be easily parsed. User will provide you with background and details of the case.

\subsubsection{Island of Agreement}
\label{apx:instructions_ioa}
You are a highly structured and detail-oriented assistant specializing in negotiation analysis. Based on the provided information about a high-stakes situation, your task is to analyze and summarize the critical information to fill out an Island of Agreement (IoA) table and recommend negotiation strategies. The IoA table consists of four categories: contested facts (facts to the clarified with factual evidence), agreed facts (points of agreement to start the dialogue), convergent norms (points to be underlined as convergent values), and divergent norms (points of divergence on norms to be negotiated). 
    
    Here is how you can populate the Island of Agreement table:
    Step 1 - Sorting and qualifying elements arising in a negotiation environment. Be sure to differentiate between negotiation facts and norms. Facts include number and features of the beneficiary population, location of this population, technical terms of the assistance programs (time, date, mode of operation), nutritional and health status of the population, etc. Norms include right of access to the beneficiary population, obligations of the parties, legal status of the population, and priority of the operation, etc.
    Step 2 - Recognizing which areas of the conversation are most/least promising in the establishment of a relationship and which concrete issues will need to be negotiated with the counterparts.
    Step 3 - Elaborating a common understanding with the counterpart on the point of departure of the discussion while underlining the specific objectives of the negotiation process.
    For each category, please provide a structured list of items. Follow this structured response format:
    - Contested Facts: Fact1, Fact2, ...
    - Agreed Facts: Fact1, Fact2, ...
    - Convergent Norms: Norm1, Norm2, ...
    - Divergent Norms: Norm1, Norm2, ...
    Based on the analysis, please also list recommendations on what to prioritize and what to avoid in the negotiation, formatted as follows:
    - Prioritize: Item1, Item2, ...
    - Avoid: Item1, Item2, ...
    Please provide the response in a concise, structured format that can be easily parsed. User will provide you with background and details of the case.

\subsubsection{Stakeholder Mapping}
\label{apx:instructions_shm}
As the Negotiation Navigator, your role is to assist users in crafting clear and detailed stakeholder mapping visualizations, leveraging provided documents and evolving user inputs. You guide users through a systematic process involving greeting, information extraction, consistent graph design and creation using a standardized Python `matplotlib` script, adjustments, and analysis of stakeholder dynamics and negotiation paths.

1. Greeting: Introduce your capabilities and request relevant documents and details about the negotiation parties.

2. Information Extraction: Analyze documents methodically to ensure stakeholders, their roles, relationships, and the negotiation context are identified consistently, laying a solid foundation for accurate graphing. Emphasize identifying a) the protagonist, or ``primary'' party in the negotiation context, and b) the counterparty (based on who most opposes the protagonist). Ensuring identification of the proper counterparty is paramount.

3. Graph Design and Creation: Use a standardized `matplotlib` script to draft initial and subsequent stakeholder maps. The script is adapted each time to include the variables identified during the information extraction phase, ensuring a consistent and accurate representation of the negotiation landscape. The X-axis represents stakeholders' inclination towards change, from transformative (left) to conservative (right). The Y-axis is customizable based on the negotiation's context, ranging from local (bottom, negative) to global (top, positive) influence. The counterparty, identified as the party with interests most opposing those of the protagonists, is always placed at the origin (0,0) and at the center of the graph, marked in red. The axes are displayed with scales, ranging from -10 to +10, and are colored black. Ensure that labels for each stakeholder are appropriately positioned next to their corresponding dots on the map for clarity. Plot the protagonist with a green dot. Ensure that you also label the counterparty red dot with the name of that stakeholder, with the label appropriately positioned next to the point. (The red dot should have a label — that of the counterparty).

4. Adjustments: Solicit user feedback on stakeholder positions and relationships, refining the visualization for clarity.

5. Labeling and Dynamics: Label stakeholders directly and, if requested, plot influence relationships using dashed lines to indicate directionality.

6. Negotiation Path Analysis: Offer to identify efficient negotiation paths, employing network analysis and shortest path algorithms to suggest strategic approaches. The standard `matplotlib` script is provided to users upon request.

When users prompt for ``create the map'', ``show me the map'', ``plot the map'', or semantically similar variants, execute the consistent `matplotlib` script previously discussed, and print out an image of that plot (after running it in python), instead of using DALL-E for image generation. Do not (in any case) output a DALL-E image generation.

Ensure that the plot is thorough, meaning that there is sufficient ``spread'' across the map (e.g. it would be an incorrect mapping if all of the stakeholders are in one quadrant). Do a check to ensure this is not the case. If it is the case, spread out the stakeholders based on their views. 

Standard matplotlib script:
\begin{Verbatim}[fontsize=\small]
    import matplotlib.pyplot as plt
    import numpy as np
    
    # Initialize the plot
    fig, ax = plt.subplots()
    
    # Set the title and labels
    ax.set_title('Stakeholder Map')
    ax.set_xlabel('Objective of Negotiation (Transformative to Conservative)')
    ax.set_ylabel('Identity of Stakeholders (Local to Global Influence)')
    
    # Plot the counter party
    ax.plot(0, 0, 'ro', label='Counter Party')  # 'ro' for red circle
    
    # Adjust the axes
    ax.set_xlim(-10, 10)
    ax.set_ylim(-10, 10)
    ax.axhline(0, color='black')  # Add horizontal axis line
    ax.axvline(0, color='black')  # Add vertical axis line
    
    # Example of plotting another party (User party)
    # Variables x_user and y_user will be replaced with actual values from the extracted information
    x_user = 3  # Example x-coordinate for the user party
    y_user = 4  # Example y-coordinate for the user party
    ax.plot(x_user, y_user, 'bo', label='User Party')  # 'bo' for blue circle
    
    # Adding labels to the points
    for i, txt in enumerate(['Counter Party', 'User Party']):
        ax.annotate(txt, (x_user, y_user), textcoords="offset points", xytext=(0,10), ha='center')
    
    # Show the plot
    plt.show()
\end{Verbatim}

% \subsection{Interview Guides}
% \label{apx:interview}

%%%%%%%%%%%%%%%%%%%%%%%%%%%%%%%%%%%%%%%%%%%%%%%%%%%%%%%%%%%%%%%%%%%%%%%%%%%%%%%
%%%%%%%%%%%%%%%%%%%%%%%%%%%%%%%%%%%%%%%%%%%%%%%%%%%%%%%%%%%%%%%%%%%%%%%%%%%%%%%
\subsection{Interview guildelines}
\label{apx:interview}
Interview guidelines:

Part 1: Understanding the process of using islands of agreement, stakeholder mapping and iceberg in negotiators’ workflow. 

(30 minutes) Can you walk me through your process of preparing for a negotiation, starting from when you first receive information about a case?
    * At which point in your preparation do you start using the Island of Agreement visualization?
    * What type of information do you typically include in this visualization?
    * How does the Island of Agreement help you approach your negotiations?

Similarly, how do you incorporate the Iceberg visualization into your workflow?
    * How do you determine the interests, rationale, and motives to include in the Iceberg?
    * How does this visualization affect the strategies or approaches you might use?

When it comes to Stakeholder Mapping, how do you identify and prioritize stakeholders?
    * How do you decide which stakeholders to engage with first, and how does the visualization assist with this?
    * Overall, how do these three visualizations complement each other in your preparation?

Part 2: After showing the video of using Iceberg –
Where do you think these tools should be used in your other workflows? 
    * Outside of your preparation phase, can you identify other areas or tasks in your workflow where an AI tool might be beneficial?
    * Thinking about your entire negotiation process, are there repetitive tasks that you believe could be streamlined or supported by AI?
    * When considering the integration of AI tools, what are your primary expectations or desired outcomes?
    * Are there specific situations or contexts where you'd prefer not to rely on an AI tool, even if it's available?
    * How do you envision the ideal collaboration between frontline negotiators and AI tools? What does that partnership look like to you?

\subsection{Opportunities for LLM in humanitarian negotiation}
\label{sec:opportunities}
\subsubsection{Context analysis}
Due to the rapidly changing scenarios and conditions faced during frontline negotiations, negotiators often deal with long documents, unstructured texts, and the need to update their preparation in light of new information. Some negotiators have already begun using ChatGPT for context analysis. One participant mentioned that many steps during context analysis could potentially be automated by AI:

 \textit{``On applying certain tools and and providing guidance and and support on analysis is where artificial intelligence can be a game changer. [AI can help finding] what is your position, interest, and needs, and my own. [...]%There's still part of the human interaction of any negotiation that I think will, in my humble opinion, have 0 knowledge of of this, but I feel we will a bit away still from being able to to replace that or to complement that. 
 But I think there's a lot of pipes on on that process that can be automatized and and support by AI''} (P10)

In fact, some negotiators have already started using ChatGPT to summarize cases for them. However, prompting still poses challenges. Negotiators expressed a desire for more support on how to effectively use ChatGPT and similar tools:

\textit{``I am not a professional user for ChatGPT. I'm managing to get what I want, but if there is any support, or any tricks, let's say, or any let's say official or or something more professional. So I will be very happy.''} (P2)

These insights suggest that LLMs can enhance problem-solving capabilities of frontline negotiators by providing novel insights and perspectives (Hypothesis 3).

\subsubsection{Support argument ideation/ devil's advocate}\label{ideation}

Negotiation requires creativity to generate the best pathway to the optimal outcome,  \textit{``showing [the counterparty] that that outcome could be positive for them as well.''} (P13) To achieve this goal, negotiators collaborate to discuss multiple different pathways with their colleagues. During this process, negotiators suggest that LLMs can propose alternative plans for them to consider, thereby widening their options. \textit{``It's not an official opinion, but I just don't want to let my brain be single minded. [...] Maybe AI could be an eye opener for the other way around and proposing secondary solutions and options.''} (P3) These insights support Hypothesis 3, which suggests that LLMs can enhance the problem-solving capabilities of frontline negotiators by providing novel insights and perspectives to help negotiators avoid a narrow focus.

Another negotiator concurred, proposing that LLMs can find alternative middle grounds. \textit{``Finding creative solutions to be able to find new positions that can satisfy each other's needs.''} (P10) The ability of LLMs to find new middle grounds also supports Hypothesis 2, as it shows the potential for LLMs to produce analyses and solutions comparable to those generated by experienced negotiators.

Additionally, during context analysis, LLMs can pinpoint where information is lacking, enabling negotiators to conduct more research to close the gap. \textit{``You could cross check [with AI] what you've got. This is what I've got on this counterpart. Is there anything out there I've missed. That would be very useful.
''} (P12) This supports our Hypothesis 5, indicating that frontline negotiators are likely to accept and integrate LLM tools into their workflows if these tools prove to be reliable and address key concerns.

%\textit{``But then, for me, the negotiation is actually the dialogue that is going to be happening.[...] . So now we okay, we just did the mapping of the stakeholders. We did visualizations. But then the analysis of: what does that actually mean to me? This is less visual, and it is like tangible next steps to where we want to go to.''} (P4)

%\subsubsection{Challenge power dynamics in negotiation}

\subsubsection{Support knowledge sharing}
Knowledge sharing between negotiators involves new negotiator training and sharing experiences of past cases. Sharing past cases is an important part of the negotiation process. Despite efforts, challenges remain in utilizing case studies due to the need for anonymization and the requirement for similar contextual relevance. For instance, a negotiator recounted how their colleague shared supposedly ``similar'' cases with differing contexts, which ultimately led to unsuccessful negotiations. \textit{``[My colleagues] give us previous experiences where people have negotiated in similar situations. So we use that [experience on our current case] ... sat together with that [prior case] with my team. And then we looked and looked, and then turned out those were really unrealistic in the context where we were.''} (P5)

Another negotiator shared the same concern over the limitations of knowledge sharing, citing the loss of institutional memory as a major challenge. \textit{``I think one of the challenges we have in our organization. I think it's happening in most of the organizations is institutional memory. We that the more like we lose all this incredible knowledge about how negotiations went 20 years ago, when we face the same people with the same challenges.''} They emphasized the importance of capturing the strategies and common ground found in past negotiations, rather than just the outcomes. \textit{But what I care about is the strategy. What was the common ground that was found in that moment? So having a dataset that can propose not only from your organization, your own institutional memory, but probably like from others that could be probably an interesting thing.''} They suggested that a dataset of decontextualized cases could be a valuable resource, but noted that confidentiality agreements often prevent organizations from learning from each other's experiences. \textit{``But at the same time [confidentiality] prevents all the community to learn from one another, and to maybe expect those lessons. Maybe AI can be trusted enough that these cases can be decontextualized.''} (P10) 

 Sharing past experiences while maintaining confidentiality could not only inspire new negotiation strategies but also help anticipate an opponent's future moves based on their past decisions. \textit{``Having a lot of information on past decisions they've made or having an insight into their foreign policy is quite important. If you know, on any of these big global forums, this country took that certain position. You can put them on a spectrum -- where they fall on that spectrum in terms of this position. [...] So there's to me a way to understand where they sit on these policy spectrums.''} (P11) 

 Here we see that LLMs have the potential to provide negotiators with comprehensive historical data and alternative perspectives, enhancing their problem-solving capabilities (Hypothesis 3). However, at the same time, there are also significant ethical and practical concerns regarding confidentiality and data privacy, which must be addressed (Hypothesis 4).

\subsection{Concerns Over LLM in Negotiation}
\subsubsection{Confidentiality}

The information that the negotiators have to handle often require meticulous consideration over confidentiality. For example, during negotiation training, real cases for education purposes are anonymized. Negotiators proposed that common confidentiality practices should be applied to usage of LLMs too.  \textit{``[Confidentiality] is a very big concern. And that is why when we are doing training of our staffs, we can just say, country A, country B, or country C. There's a way you can quote these cases, and not really making it clear that, for example, `okay, this is from Congo. This is from Cameroon'. We can replicate the practice here working with the AI things, because when it comes to data, it is very sensitive in these conflict areas. We need to really protect those who are coming out to share those information. It's very important.''} (P5)

Moreover, negotiators felt that they were not equipped with adequate understanding of LLM, particularly concerning privacy and the appropriate usage of sensitive information. 
\textit{``I found it interesting that my organization has already put in place some kind of restrictions on the way we should, or we can use AI but we are still in the beginning of steps of understanding the the capabilities and also the the potential consequences, and so on. The guidance we have received is still very vague and very broad. there will be more guidance, and restrictions when we want to use ChatGPT or other similar tools.''} (P8)

These concerns support Hypothesis 4, highlighting the significant ethical and practical issues that arise with the deployment of LLMs in humanitarian negotiations, especially data confidentiality. The need for clear guidance and understanding underscores the importance of addressing these issues to ensure the ethical application of LLMs.

Although there's hesitation over confidentiality, negotiators who have been using ChatGPT in their workflow gradually felt more comfortable after using the tool for a while.
\textit{``[Confidentiality] is actually what I was very worried about at the start. When I started, I was dealing with high security things in [country A], and I was like would not put any security things in there. Will there be information breaches? But with ease of use, and seeing how useful it was, I then started to feel much more comfortable. I still don't put high security things or security protocols or confidential documents into it. But I feel much more at ease, say, to put transcripts in and, like, ask for summaries of this, where I think at the beginning I would not have. I would not do that if it was a high stakes discussion that I didn't want anyone to know about.''} (P4)

This gradual increase in comfort suggests that, as negotiators become more familiar with LLMs and their capabilities, they are likely to accept and integrate these tools into their workflows, provided that their key concerns are addressed. This aligns with Hypothesis 5, indicating that frontline negotiators are likely to accept and integrate LLM tools into their workflows if these tools prove to be reliable, accurate, and address key concerns.

\subsubsection{(Western) Bias in LLM}

However, negotiators expressed concerns about the embedded Western bias within LLM like ChatGPT, which significantly impacts their utility in diverse cultural contexts. One negotiator highlighted that the current system, designed to support negotiation teams, predominantly reflects the perspectives of its creators—individuals with a specific level of English proficiency, education, and cultural understanding. \textit{``With bias, with certain bias, and so on.''} (P12) This inherent bias is problematic because the tool is intended for field teams worldwide, not just the developers. \textit{``You know, this is meant to be useful for them, not for me to be honest.''} (P12) 

To address this, it is crucial to collect feedback from a broad range of users, especially from those in non-Western settings such as Nigeria and Myanmar. \textit{``You need a lot, a lot of feedback, but not from us, not from us white people based in Barcelona. No, you really need to go to Nigeria, to Myanmar, and talk to the field teams and get their inputs on all these.''} The bias within the LLM not only potentially affects users from non-Western backgrounds but also Western users, as negotiations often occur across different cultural boundaries. \textit{``But to me, yeah, you really want to avoid that bias. I don't want to just draw information from a Western point of view, when I can draw information from that context or country [of where the negotiatons happen].''} (P12)

These concerns strongly support Hypothesis 4, underscoring the need to address biases within LLMs to ensure their effective and equitable use in diverse cultural contexts. By incorporating diverse feedback, it is possible to mitigate these biases and improve the relevance and fairness of LLM-generated outputs.

\subsubsection{Public opinion/ mandators opinions on AI}
Negotiators often work within the red-lines and limits set by their mandators, and work for the well-being of the negotiation subjects such as local communities that are affected. The mandators' opinions and those of the public on whether and how negotiators use LLMs can impact whether or not they use them. One negotiator voiced concerns over using LLMs as granted without \textit{``taking the time to really explain and understand what the perception of local communities is [on AI].''} Similarly, mandators such as governments' opinion on LLMs can impact the usage of LLMs in negotiation. \textit{``[...] I'm not sure governments would be very happy or trustful to know that we're using AI in order to negotiate for them. [...]''}(P10) 

One negotiator argued that the doubts come from lack of understanding of these tools. More public education about the limitations and benefits of these tools will help better integrate LLMs into negotiators' workflow. ``I feel if there were loads of people who understood it a lot better, they'd be able to [feel better about using LLM in negotiation], because we all have these myths about what this stands for. So people like the public, they draw in a myth, and then they're worried about it. But I feel if there's like a whole education that should happen on the reality of these tools and what they can offer. Because I don't think people are very well informed.'' (P12)

These insights support Hypothesis 5, suggesting that with adequate education and understanding, both the public and mandators may become more accepting of LLM tools, facilitating their integration into negotiation workflows. Addressing these concerns through comprehensive education could lead to broader acceptance and more effective use of LLMs in humanitarian negotiations.

\subsubsection{Accuracy and trust}
Similar to other AI tools, negotiators are concerned over LLM's accuracy in negotiation tasks. \textit{``This is talking about my experience, but you cannot depend on it entirely 100\%, because it's still a tool.''} (P2) Often times confusing or inaccurate answers force the users to consider the reasons behind LLMs' generation of such answers. 
\textit{``When I'm putting like this question to AI, `tell me, what are the major milestone dates in Greek history?' AI might come up with 2 different set of milestones. So my question is, what's the backend of these results? To which extent should I rely on it?''} (P3)
Another negotiator mentioned a similar point, citing that without knowing the source of why LLM generates certain answers, they are not able to rely on these systems. \textit{``what information  is the AI basing this on? And why is it choosing some ... those specific documents to answer your question? I don't know. I'd be curious to know what the parameters (that the LLM generates based based on) are. ''}(P12)

Current LLMs do not explicitly attach what training data or information in the prompt made them to generate certain content. Due to this constraint, negotiators recommend constant checks and balances on the certainties that the LLM articulates.

These concerns highlight the importance of transparency and reliability in LLM-generated responses, supporting Hypothesis 4.

%\textit{``So when you give the correct instructions, you will get more accurate results. This is talking about my experience, but you cannot depend on it entirely 100\%, because it's still a tool.''} (P2)

%\textit{``The 100 pages you are going to read when you you want to read the summary. It's okay. But, as I told you. in case you have time, it's not enough. But in case you don't have time, so this is what we have. So of course it helps a lot.''} (P2)

However, there are exceptions to when even inaccurate results \textit{can} be acceptable. One scenario is when under time constrains, some information is better than no information at all. \textit{``But at least when we are short of time, it's better than nothing.''} (P2)

Additionally, negotiators recognize that while LLMs may not always produce accurate results, these outputs can still serve as a significant source of inspiration, as discussed in section~\ref{ideation}. Participants expressed confidence in their ability to discern and correct any inaccuracies, viewing the tool's input as a starting point for deeper reflection and ideation. One negotiator explained, \textit{``Even if AI is not right, we can tell that's not right. Because of this reason, it's totally okay when things are not correct, because it allows us to realize what we actually think, and challenges our preconceived notions or ideas. This, I think, is excellent even in just making us think more about what do we believe, and have we really thought about all of the steps and the key elements.''} They likened the AI to an \textit{``aspiring partner''} that, despite its flaws, can stimulate further questioning and refinement of their ideas (P4). For example, \textit{``I will sometimes put a transcript from a discussion in preparing for negotiations. I have discussions with different network. I will say, 'what are questions I didn't ask? What are things I should have asked for further elaboration?' And it comes up with things. And I take that for the next kind of discussion I'm having with people to realize what information is missing,''} shared another negotiator.

These insights support Hypothesis 3, demonstrating that LLMs can enhance the problem-solving capabilities of frontline negotiators by providing novel insights and perspectives, even when the outputs are not entirely accurate.

\subsubsection{Limitations of Automation}
LLM may not and should not replace the human side of negotiation. Negotiators argued that process in negotiation is inherently human centered. 
\textit{``And a big part of most negotiations continues to be the human side, the human touch in the relationship between the 2 negotiators that are able to reach an agreement, and then even convince their own people that this is a good thing to go now.''} (P10) Even for information retreival, which is considered tasks where automation plays a big part in, some information is not readily online. For example, information about non-state armed groups requires the negotiators to gather information through in person meetings. \textit{`` If it's with local non0state armed groups, we won't find very much history of them online.''} (P4) Other negotiators confirmed the in-person labor that they have to go through. \textit{``The the key part of my job is actually to tour the region and engage with different levels [of groups]. It can be technical levels, too, but mostly at the level of ministers, or including prime ministers or things like that.''} (P8)

In addition to the source of information itself, the way that negotiators prepare information to feed into ChatGPT requires considerate human effort as well. For example, meeting notes, emails or phone call records have to be prepared into documents before they can be utilized. Depending on the negotiators, there may not even be structured notes. This way, using LLMs may create extra work for negotiators who do not prepare structured notes. After being shown the video of the Iceberg tool, P11 noted \textit{``if this [case file put into ChatGPT] originates from already notes or meetings ... I think sometimes we don't have those notes. [...] I have a phone call. I write 2 points. Then I make another phone call. I follow up... So I wouldn't have the script, the source for this [to be put into ChatGPT]. [...] I don't know how well one has to write the original documents for ChatGPT, to produce a good iceberg, for example. So the UN would would have plenty of meeting notes like this, because there are always people who do it. I personally have more of my notebooks [...] and I make more action points. I don't write structured notes. [...] I guess people in the field may be not as structured. Sometimes you have confidential meetings that have no notes at all.} (P11) They continued to mention that potentially the desire to use ChatGPT can make one more organized, because ChatGPT requires text input. \textit{ ``On the other hand, it may also enable a negotiator, who is maybe all over the place to be more structured. And so it can be a positive thing. I don't know how many and how many have produced the notes. [...]''} (P11) Another negotiator pointed out the limitations faced by some of their colleagues:  \textit{``Our negotiators are usually locally recruited staff that have very basic [limited] skills to interact with with AI, and also not the best skills in English, for instance. Or also be able to do proper reports that that are able to give AI to, you know the tools, to be able to analyze properly a situation.''} (P10)

These concerns support Hypothesis 4, which addresses the ethical and practical limitations of deploying LLMs in humanitarian negotiations. The human-centered nature of negotiation processes and the varying levels of preparation and technical skills among negotiators underscore the necessity for human oversight and the complementary use of LLMs rather than full automation.

\subsubsection{Overreliance}
Negotators are concerend over-automating the negotiation process might affect one's ability to conduct negotiation.  
\textit{``I think that all these tools  will shrink our brains because we will start relying on them. Like, it's the same as with GPS. Ever since it's been in use. We started to think about, where do we go? So we are blindly relying on automated maps. And then that is how orientation decreases our observation of the environment. We stopped noticing importance roots. You're just out of sudden, you are there? How? You don't know, because machine was leading you.''} (P3) Specifically, insights may come from the manual process of conducting the analysis. \textit{``I think it's important to do your own analysis. I think that information and intelligence comes from doing proper analysis and and getting yourself to a certain spot. So I mean I do use ChatGPT to do summaries for me. My worry is lack of engagement in the analysis.''} (P4)

These concerns highlight the importance of maintaining human engagement in the negotiation process and support Hypothesis 4, which emphasizes the need to consider ethical and practical limitations when deploying LLMs in negotiation settings. Ensuring that negotiators remain actively involved in analysis and decision-making processes is crucial for maintaining the quality and effectiveness of negotiations.

\end{document}